\documentstyle[aps,prl,epsf,twocolumn]{revtex}

\author{
Hans-J\"urgen Sommers$^1$,
Dmitry V. Savin$^{1,2}$, and
Valentin V. Sokolov$^{2}$ }
\address{ $^1$Fachbereich Physik, Universit\"at-GH Essen,
45117 Essen, Germany}
\address{$^2$Budker Institute of Nuclear Physics,
630090 Novosibirsk, Russia}

\title{Distribution of proper delay times in quantum chaotic scattering: \\  
A crossover from ideal to weak coupling}
\date{Received 8 May 2001; published 8 August 2001:
Phys. Rev. Lett. {\bf 87}, 094101 (2001)}

\begin{document}
\draft

\twocolumn[
\widetext
\begin{@twocolumnfalse}
\maketitle

\begin{abstract}
The probability distribution of the proper delay times during scattering on
a chaotic system is derived in the framework of the random matrix approach
and the supersymmetry method. The result obtained is valid for an arbitrary
number of scattering channels as well as arbitrary coupling to the
energy continuum. The case of statistically equivalent channels is studied
in detail. In particular, the semiclassical limit of infinite number of
weak channels is paid appreciable attention.
\end{abstract}
\pacs{PACS numbers: 05.45.Mt, 03.65.Nk, 24.60.-k, 73.23.-b}
%
%

\vspace{0.5cm}
\narrowtext
\end{@twocolumnfalse}
]

\narrowtext  

The temporal aspect of collision and transport phenomena has, starting with
the pioneering works \cite{Wig55,Smith60}, repeatedly attracted attention.
Analysis of the  duration of a process gives an interesting and important
information complementary  to that delivered by the energy representation.
The temporal approach becomes especially  elucidating in the
cases of chaotic resonance scattering and transport in disordered media
when many complicated long-lived intermediate states are involved
\cite{Lyub77,LW91,HDM91,Eckh93,ISS94,LSSS95a,GMB96,FS97,FSS97,BFB97,SS97,KS00,GKK99,SFS01}; see \cite{FS97,Dittes00} for recent reviews.

The delay  of an almost monochromatic wave packet during an $M$-channel
process is conventionally described by the $M$$\times$$M$ Wigner-Smith matrix
$Q$=$-i\hbar S^{\dagger}\partial S/\partial E$, with $S$ being the
scattering matrix at the collision energy $E$. In the resonance scattering,
the matrix  element $Q_{cc'}$ describes the overlap of the internal parts of
the scattering wave functions in the  incident channels $c$ and $c'$
\cite{SZ97}. In particular, the diagonal element $Q_{cc}$, which  is
interpreted as the mean time delay in the channel $c$, gives the norm of the internal part $b^{c}(E)$ of the scattering wave function excited through the channel $c$ (the normalization is fixed as  usual by setting the flow in any given entrance channel to one). This directly relates the Wigner-Smith matrix to the effective non-Hermitian Hamiltonian ${\cal H}$=$ H \!-\! (i/2)VV^{\dagger}$ of the unstable intermediate system as follows (henceforth $\hbar$=1) \cite{SZ97}:
\begin{equation}\label{Q}
Q(E) = V^{\dagger}\frac{1}{(E-{\cal H})^{\dagger}}\,
     \frac{1}{E-{\cal H}}V  \equiv  b^{\dagger}(E)\, b(E)\,.
\end{equation}
The Hermitian part $H$  determines the appropriate basis for the internal
motion whilst the amplitudes $V_n^c$ describe the coupling between $N$
interior and $M$ channel states.
We adopt below the standard statistical approach \cite{VWZ85} based on the
random-matrix theory (RMT) to simulate the complicated intrinsic motion; see
\cite{BeenRev,GMGW98}.

Recently, an appreciable success has been achieved by the authors of
Refs.\cite{BFB97}. By quite a distinguished way of reasoning, they managed to
calculate in the  framework of RMT the joint probability distribution of the
eigenvalues $q_c$ of the Wigner-Smith matrix, which are called the {\it
proper} delay times. Using then the method of orthogonal  polynomials they
calculated also the density of proper delay times. These findings received
later experimentally relevant applications to quantum dots \cite{dots}
and optics of random media \cite{optics}. However, all the results are
restricted to the case of perfect coupling to the continuum (all transmission
coefficients are equal to 1).

In this Letter we show that representation (\ref{Q}) allows us to
extend at least part of the results mentioned above to the case of arbitrary
coupling. Namely, we calculate the probability distribution of the proper delay times $q_c$
\begin{equation}  \label{P}
{\cal P}_M(q ) = \frac{1}{M}\,\langle\sum^M_{c=1}\delta(q-q_c)\rangle
               = \frac{1}{\pi}\,{\rm Im}\,G(q -i0)\,,
\end{equation}
where $\langle\ldots\rangle$ denotes the statistical average, and
\begin{equation}\label{Green}
G(z) = M^{-1} \,\langle {\rm tr\,} (z - Q)^{-1}\rangle
\end{equation}
is the trace of the Green's function of the Wigner-Smith matrix. The positive definite form (\ref{Q}) of  $Q$ provides the analyticity of  $G(z)$  throughout the complex $z$-plane save the discontinuity line along the positive part of the real axis $q$=Re$\,z\!>\!0$. Function (\ref{P}) is normalized to unity.

The function $G(z)$ can be readily obtained from the generating function
($z_{\pm}$=$z{\pm}\omega/2$)
\begin{equation} \label{Z}
Z(z,\omega) = \left\langle \frac{\det (z_+ -Q)}{\det (z_- -Q)}\right\rangle
 = \left(\frac{z_+}{z_-}\right)^M
   \left\langle \frac{{\rm Det} (A_+)}{{\rm Det} (A_-)} \right\rangle
\end{equation}
as $G(z)=M^{-1}{\partial Z(z,\omega)}/{\partial \omega}|_{\omega=0}$.
We have exploited here the explicit form (\ref{Q}) to cast $Z(z,\omega)$ in
terms of determinants of $N$$\times$$N$ matrices $A_{\pm}$ (rather than those
of $M$$\times$$M$ ones $z_{\pm}-Q$). Now, we use the identity
\begin{eqnarray}
A_{\pm}
&\equiv& (E-{\cal H})(E-{\cal H})^{\dagger} - VV^{\dagger}/z_{\pm} \nonumber\\
&=& (E-{\cal H}-i/z_{\pm}) (E-{\cal H}^{\dagger}+i/z_{\pm}) - 1/z_{\pm}^2
\end{eqnarray}
and double the dimension, introducing Pauli matrices $\sigma_{i}$ in the
doubling space. As a result, the determinants get the following simple form:
\begin{equation}
{\rm Det}(A_{\pm}) = {\rm Det} \bigl( \frac{1}{2}VV^{\dagger} -
\frac{1}{z_{\pm}}(1 \!+\! \sigma_1) - i(E \!-\! H)\,\sigma_3 \bigr),
\end{equation}
which allows us to represent the generating function (\ref{Z}) in form of a
Gaussian integral over an auxiliary supervector and therefore to
employ hereafter the standard supersymmetry technique \cite{VWZ85,Efetov}.
As a result, we come to the following (exact at $N\!\to\!\infty$) expression
for the generating function
\begin{eqnarray}\label{Zsaddle}
Z(z,\omega) &=& \left(\frac{z_+}{z_-}\right)^M \!\! \int [d\hat\sigma]
\exp\left[ \frac{\pi}{2\Delta} {\rm str}(U(\Lambda-i\Lambda_1)\hat\sigma) \right]
\nonumber\\
&& \times  \prod_{c=1}^M
{\rm sdet}[1+\gamma_c\Lambda(iE/2+\pi\nu(E)\hat\sigma)]^{-1/2} \,
\end{eqnarray}
as the integral over a noncompact saddle-point manifold
$\hat\sigma$. The 8$\times$8 supermatrices
$\Lambda$, $\Lambda_1$ appearing above
are the supermatrix analog of the Pauli matrices $\sigma_3$, $\sigma_1$, and
$U$=${\rm diag}(z_-^{-1}{\bf 1}_2,z_+^{-1}{\bf 1}_2,
z_-^{-1}{\bf 1}_2,z_+^{-1}{\bf 1}_2)$.
Definitions of the superalgebra as well as the explicit parameterization of
$\hat\sigma$ and the invariant measure can be found in \cite{VWZ85,Efetov}.
The average density of states $\nu(E)$=$\pi^{-1}\sqrt{1\!-\!(E/2)^2}$
determines the mean level spacing $\Delta$=$(\nu N)^{-1}$ of the closed
system. Phenomenological constants
$\gamma_c$$ > $0 originate from the coupling amplitudes:
$(V^{\dagger}V)_{cc'}$=$2\gamma_c\delta_{cc'}$; they enter final expressions
only by means of the transmission coefficients
$T_c$=$1\!-\!|\overline{S}_{cc}|^2$=2$[1
\!+\!(\gamma_c\!+\!\gamma^{-1}_c)/2\pi \nu(E)]^{-1}\!\leq\! 1$ \cite{VWZ85}.
The latter are the conventional characteristics of coupling of an unstable
system to the energy continuum.

Suspending all technical details of quite standard calculations [for the
sake of simplicity, we restricted them to the unitary symmetry class which
corresponds to  systems with broken time-reversal symmetry (TRS)] to a more
extended publication, we proceed with the following result for the Green
function:
\begin{eqnarray}\label{G(z)}
G(\zeta) &=& \frac{1}{\zeta} +\frac{1}{M\zeta^2} +
\frac{1}{2M\zeta} \int^{\infty}_{1}\!\! d\lambda_1
\!\int^{1}_{-1}\!\frac{d\lambda_2}{\lambda_1 \!-\! \lambda_2}
\prod^{M}_{c=1}\frac{g_c \!+\! \lambda_2}{g_c \!+\! \lambda_1} \nonumber \\
&&\times
\left( f(\lambda_2)\frac{\partial b(\lambda_1)}{\partial\zeta} -
b(\lambda_1)\frac{\partial f(\lambda_2)}{\partial\zeta}\right)\,.
\end{eqnarray}
Henceforth, we scale the original variable $z$=$\zeta t_H$
[and correspondingly the Green's function (\ref{Green})]
in the natural units of the Heisenberg time $t_H$=$2\pi\hbar/\Delta$. The constants $g_c\!=\!2/T_c \!-\! 1$ are related to the transmission coefficients $T_c$. At last, we denote
$b(\lambda_1) $=$ e^{\lambda_1/\zeta}{\rm I}_0(\zeta^{-1}\sqrt{\lambda_1^2-1})$
and
$f(\lambda_2) $=$ e^{-\lambda_2/\zeta}{\rm J}_0(\zeta^{-1} \sqrt{1-\lambda_2^2})$,
with ${\rm I_0}(x)$ (${\rm J_0}(x)$) being the modified (usual) Bessel
function of zero'th order.
As it stands, the Green's function (\ref{G(z)}) is an
analytic function of the complex variable $\zeta$ in the half-plane 
${\rm Re}\,\zeta\!<\!0$. One can deform the original $\lambda_1$-integration to
that along the imaginary axis [1,+$i\infty$) [or [1,-$i\infty$)] for
Im($\zeta^{-1}$)$>$0 [or Im($\zeta^{-1}$)$<$0], so that $G(\zeta)$ is
analytically continued to the whole complex $\zeta$-plane with a cut along
positive Re$\,\zeta$. Applying (\ref{P}), we get finally the following expression for the probability distribution of the proper delay times 
\begin{equation}\label{P(t)}
{\cal P}_M (t\!=\!q/t_H) = \frac{1}{Mt} \sum^M_{c=1} \left(
F_c^F \frac{\partial F_c^B}{\partial t} -
F_c^B\frac{\partial F_c^F}{\partial t} \right)\,,
\end{equation}
for $t$$>$0, and ${\cal P}_{M}(t)\!\equiv\! 0$ for negative times.
Here
\begin{eqnarray}
&& F_c^B = e^{-g_c/t}\, {\rm I_0}(t^{-1}\sqrt{g_c^2 \!-\! 1})
    \prod_{a (\ne c)}{1\over g_a \!-\! g_c}            \label{FB}\,, \\
&& F_c^F = \frac{1}{2}\!\int_{-1}^{+1}\!\! d\lambda \, e^{-\lambda/t}\,
   {\rm J_0}(t^{-1}\sqrt{1 \!-\! {\lambda}^2})
   \prod_{a (\ne c)}{( g_a \!+\! \lambda )}\,.             \label{FF}
\end{eqnarray}
The function (\ref{FF}) is, in fact, a polynomial in $1/t$.
The formulae (\ref{P(t)})--(\ref{FF}) are valid for an arbitrary number of
scattering channels $M$ and arbitrary constants $g_c \!\geq\! 1$. In the
situation when all chaotic states of the target system are alike, the
channels get statistically equivalent.  Setting all $g_c$=$g$, we
arrive in such a case at
\begin{equation}\label{Peq}
{\cal P}^{\rm eq}_M(t) = \frac{1}{Mt} \sum^{M-1}_{M'=0}
  \left(F_{M'} \frac{\partial B_{M'}}{\partial t} -
  B_{M'}\frac{\partial F_{M'}}{\partial t} \right)
\end{equation}
with
\begin{eqnarray}
B_M &=& \frac{1}{M!} (\frac{-\partial}{\partial g})^M
   e^{-g/t}\, {\rm I_0}(t^{-1}\sqrt{g^2 \!-\! 1})  \,, \label{BM} \\
F_M &=& \frac{1}{2}\int_{-1}^{+1}\!d\lambda \,e^{-\lambda/t}
   {\rm J_0}(t^{-1}\sqrt{1 \!-\! {\lambda}^2})
   \,(g \!+\! \lambda )^M                                \nonumber    \\
    &=& \sum_{m=0}^M \frac{1}{(2m+1)!}\left(\frac{\partial^2}{\partial g^2}-
\frac{2}{t}\,\frac{\partial}{\partial g}\right)^m g^M\,. \label{FM}
\end{eqnarray}
It is worth noting  that, due to similar analytical structure of the
functions $F_c^{B}$ and $F_c^{F}$, distribution (\ref{P(t)}) decays as $t^{-M-2}$ in the limit $t$$\rightarrow$$\infty$.
This is in agreement with the known universal law $t^{-\beta M/2-2}$
\cite{LW91,FS97,FSS97,BFB97,SS97} valid for all Dyson's symmetry classes
$\beta$=$1,2,4$ ($\beta$=2 in the present case of broken TRS).

In the single-channel case, $M$=1, Eqs.~(\ref{P(t)}) and (\ref{Peq})
confirm the result  ${\cal P}_1(t)$=$t^{-1}\partial B_0/\partial t$
found earlier in \cite{FS97}. 
When the constant $g\!\approx\!2/T\!\gg\!1$ this distribution shows a narrow maximum at  the small time $t\!\sim\!(2g)^{-1}\!\ll\!1$
because of the cooperative influence of the tails of many remote resonances.
The far asymptotics ${\cal P}_1(t)\!\propto\! t^{-3}$, $t\!\gg\! g$, is due to
fluctuations of the widths of many narrow resonances
\cite{HDM91,ISS94,Dittes00}. At last, in the parametrically wide
domain $(2g)^{-1} \!\ll\! t \!\ll\! g$ the  distribution decreases as $t^{-3/2}\,$ \cite{FS97,FSS97}. Just this domain saturates in the case $g\gg 1$ the well known \cite{Lyub77} sum rule 
$\langle t\rangle$ =$\langle {\rm tr}Q\rangle$/$t_H$=$1$, which is, actually, satisfied regardless of the coupling strength to the continuum [see Eq.~(\ref{trQ}) below].

At the same time, the average time spent
by a spatially small wave packet in the interaction region ({\it collision} time) is equal to 
$\langle\tau\rangle \!=\! \langle Q\rangle/T \!=\! t_H\langle t\rangle/T$ \cite{Lyub77,HDM91,LSSS95a,Dittes00}. 
This relation suggests  splitting the transmission coefficient off the delay time of an almost monochromatic wave: $q$=$T\tau$.
The new variable $\tau$ is then the time the projectile is delayed inside the interaction region after having penetrated there with the probability $T$. In agreement with the above consideration, the most probable duration $\tau$ of
the stay inside is given by the Heisenberg time $t_H$.

In the case of an arbitrary number of channels $M$ an additional scaling $t_s$=$Mt$ provides  a similar condition  
\begin{equation}\label{trQ}
\langle t_s = qM / t_H=\Gamma\,\tau \rangle
= \int^{\infty}_{0}\! dt_s \,t_s  {\cal P}_s(t_s) = 1 \,.
\end{equation}
 Here ${\cal P}_s(t_s)$$\equiv$
$M^{-1}{\cal P}_M(t$=$t_s/M)$ and the mean transmission coefficient
${\overline T}$=$M^{-1}\sum_c T_c$ has been set off, $q$=${\overline T}\tau$.
The scaled variable $t_s$ measures the time $\tau$ in units of a
typical life time. Indeed, the characteristic width of a state which decays
through $M$ open channels is estimated as $\Gamma$=$\sum_c T_c\,\Delta/2\pi$
\cite{LW91}.
\begin{figure} \label{fig1}
\unitlength 1cm
\begin{picture}(8,5)
\epsfxsize 8cm
\put(-0.1,0){\epsfbox{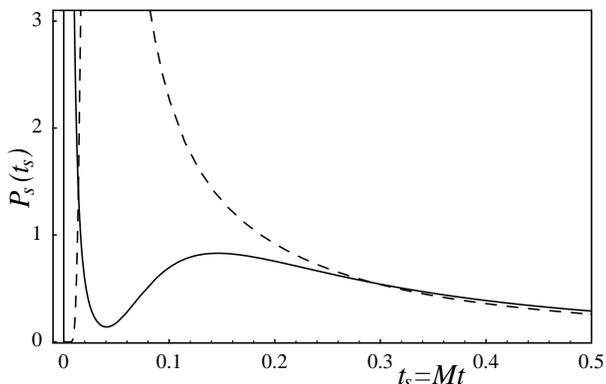} }
\end{picture}
\caption{ The distribution ${\cal P}_s(t_s)$ of the scaled proper delay
times at $M$=2 and ${\overline T}$=0.25 for the case of equivalent, $\delta
T$=0, and strongly nonequivalent channels, $\delta T$=0.49, (dashed  and
solid lines, respectively). }
\end{figure}
Figure 1 demonstrates, using the example of the two-channel distribution
${\cal P}_2(t)$, the dependence on the strengths of individual channels. To
fetch out the influence of channel nonequivalence, we keep the
openness characterized by $T_1\!+\!T_2$=2${\overline T}$ fixed and change
the  difference $|T_1\!-\!T_2|$=$\delta T$. Two distinct time scales with
similar probabilities are well seen when $\delta T$ is close enough to its maximal value min(2${\overline T}$,$2\!-\!2{\overline T}$). Fast  processes appear when one of the transmission coefficients becomes small.

We consider now in detail the case of large number of statistically
equivalent channels, $M$$\gg$1. The regime of isolated resonances, $\Gamma
\!\ll\! \Delta$, corresponds to the condition $g \!\gg\! M \! \gg$1. In this
limit the widths of resonances become statistically  independent and
expression (\ref{Peq}) reduces to the probability distribution of the
{\it partial} delay times investigated in much detail in \cite{FS97},
see also \cite{FSS97,SFS01}.

Here we focus our attention rather on the semiclassical limit of strongly
overlapping resonances excited through and decaying into a number $M$ of
channels, which is scaled with $N$, the strength constant $g$
being kept arbitrary. Similar to Refs. \cite{LSSS95b,LSSS95a},  an
additional saddle-point approximation can be used in this case to perform the
integration  in (\ref{Zsaddle}).
Under the physically justified condition \cite{LSSS95a} $M/N\!\ll\! 1$ (though
both $M,N$$\to$$\infty$) one finds that
$P(t_s) \!\equiv\! \lim_{M\rightarrow\infty} {\cal P}_s(t_s) \! = \!
(\pi t_s^2)^{-1}{\rm Im}\,K(t_s)$,
where the function
$K(t_s) \!\equiv\! t_s^2( G(t_s)\!-\!t_s^{-1})$ satisfies a
cubic equation of the form
\begin{equation}\label{cubic}
K(K^2+2g\,K+1) - {t_s}(K^2-1)=0\,.
\end{equation}
Being derived for the case of the unitary ensemble this result remains
actually valid for all tree symmetry classes. The calculation just described
sets a new intrinsic time scale of the problem \cite{LSSS95b}: the empty gap
$\Gamma_g$ between the real axis in the complex energy plane and the cloud
of the resonances in its lower half plane. In the limit considered $\Gamma_g$
coincides exactly with the  well-known Weisskopf width
$\Gamma_W$=$MT\Delta/2\pi$.

The case of ideal coupling, $g$=1, is especially simple since the cubic equation (\ref{cubic}) readily reduces to a quadratic one which immediately leads to the result
\begin{equation}\label{Pid}
P_{\rm id}(t_s) = (2\pi t_s^2)^{-1}
             \sqrt{(t^{\rm id}_{-} - t_s)(t_s - t^{\rm id}_{+})}\,,
\end{equation}
with $t^{\rm id}_{\pm}$=$3\!\pm\!\sqrt{8}$, obtained earlier in \cite{BFB97}
from the Laguerre ensemble. For an arbitrary value of $g$
the searched distribution is expressed as
\begin{equation}\label{Pex}
P(t_s) = \frac{\sqrt{3}}{2\pi t_s^2} \left[\left(|r|+\sqrt{D}\right)^{1/3} -
\left(|r|-\sqrt{D}\right)^{1/3}\right]
\end{equation}
in terms of two polynomials: $D$=$-3\eta^4 \!+\! 2g\eta^3 \!+\!
11\eta^2/3 \!-\! 4g\eta \!+\! g^2 \!+\! 3^{-3}$ and
$r$=$\eta^3 \!-\! 2\eta \!+\! g$,
where
$\eta \!\equiv\! (2g \!-\! t_s)/3$.
The density does not vanish in the domain where  the discriminant
$D\!>\!0$. It is obvious that the forth order polynomial $D$ always
has two real roots $\eta_- \!<\! 0$ and $\eta_+ \!>\! 0$ and $D$ is positive
only inside the region  $\eta_- \!<\! \eta \!<\! \eta_+$.
Therefore, similar to the case of perfect coupling, the time $t_s$ is
always restricted to a
finite domain $t_- \!<\!t_s \!<\! t_+$.
Near the edges $t_{\pm}$=$2g \pm 3|\eta_{\mp}|$
the discriminant $D$
is small and expression (\ref{Pex}), as in (\ref{Pid}), simplifies to
\begin{equation}\label{Pedg}
P(t_s) \approx  (\pi t_s^2)^{-1} |r|^{-2/3} \sqrt{D/3} \propto
\sqrt{|t_s-t_{\pm}|}\,.
\end{equation}

A more detailed investigation is possible in the case of week individual
channels, $g$$\gg$1. All the roots of $D$ in this case can
be found as  expansion due to the smallness of $1/g$.
Near the lower edge $t_-$$\approx$1/8$g$ Eq.~(\ref{Pedg}) gives
\begin{equation}\label{P_-}
P(t_s) \approx \frac{g}{\pi (2gt_s)^{3/2}} \sqrt{4 - (2gt_s)^{-1} }\,.
\end{equation}
In this case  the term $|r|$ dominates in Eq.~(\ref{Pex}) and therefore
 Eq.~(\ref{P_-}) is valid in the parametrically large domain
$1/8g \!\lesssim\! t_s \!\ll\! 2g \!-\! \frac{3}{2}(g/2)^{1/3}$ where
the influence of other (three) roots remains negligibly weak \cite{remark}.
When $t_s\!\gg\! 1/8g$ this distribution  displays the universal
behavior $t_s^{-3/2}$. Such a law is a typical and most robust feature of the time-delay distributions in a weakly open chaotic system and  may be expected from quite a general argumentation \cite{FS97,FSS97}. 
Within a relatively narrow domain
$-\frac{3}{2}(g/2)^{1/3} \!\lesssim\! t_s \!-\! 2g \!\ll\! 3(g/2)^{1/3}$ the
distribution develops a local minimum and maximum, which are due to 
a pair of the complex roots 
$t^{\pm}$$\approx$$2g \!-\! \frac{3}{2}(g/2)^{1/3}(1 \!\pm\! i\sqrt{3})$. At last, very close to the upper edge $t_+$$\approx$$2g \!+\! 3(g/2)^{1/3}$
the distribution sharply changes to the form
\begin{equation}\label{P_+}
P(t_s) \approx \frac{2 \,(g/2)^{1/6}}{\sqrt{3}\,\pi t_s^2}\sqrt{t_+-t_s}
\end{equation}
provided by Eq.~(\ref{Pedg}). Figure 2 illustrates these results.
\begin{figure} \label{fig2}
\unitlength 1cm
\begin{picture}(8,5)
\epsfxsize 8cm
\put(-0.1,0){\epsfbox{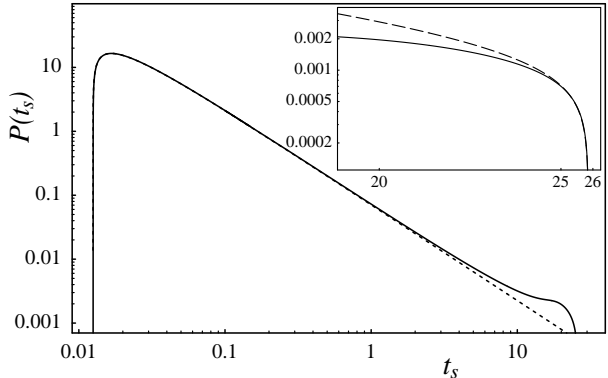}}
\end{picture}
\caption{ The distribution $P(t_s)$ in the limit $M\!\to\!\infty$. The exact
(\ref{Pex}) and approximate expressions (\ref{P_-}) (dotted) and
(\ref{P_+}) (dashed line, see inset) are shown at $g$=10. An intermediate $t_s^{-3/2}$ behavior is clearly seen.}
\end{figure}

With the help of the distribution found the Green's function can  be
represented in the spectral form
$ G(z)=\int_{t_-}^{t_+} dt_s P(t_s)/(z-t_s) $.
This function is analytical both at the origin and infinity and therefore
can be expanded near these points in power series. Such an expansion yields
the general formulae  for the negative and positive moments of the
distribution ($n \!\geq\! 0$)
$$ 
\langle t_s^{-(n+1)}\rangle \!=\!
\frac{-1}{n!}\,\frac{d^nG(z)}{dz^n}\Big|_{z=0}\,, \
\langle t_s^n\rangle \!=\!
\frac{1}{n!}\,\frac{d^n}{d\zeta^n}
\frac{G(\zeta^{-1})}{\zeta}\Big|_{\zeta=0}\,.
$$ 
All the moments are finite. The derivatives can directly be
calculated by making use of the cubic equation (\ref{cubic}).

In conclusion, we have derived the distribution of proper delay times in
chaotic scattering at arbitrary number of scattering channels and arbitrary
coupling to continua.  The appearance of distinct time scales is traced to the properties of the delay time distribution with increasing degree of channel nonequivalence. The physically interesting case of many equivalent channels weakly coupled to continua is studied in much detail.

We are grateful to Y.V. Fyodorov and P.G. Silvestrov for comments.
The financial support by
SFB 237 ``Unordnung and grosse Fluktuationen'' (H.J.S. and D.V.S.) and
RFBR Grant No. 01--02-17621 (V.V.S.) is acknowledged with thanks.


\begin{thebibliography}{99}
\bibitem{remark}
We note that (\ref{P_-}) can easily be derived from (\ref{cubic}) if one
neglects there the term $K^3$ as long as $t_s\!\ll\! g$.
\end{thebibliography}
\end{document}